\documentclass[aps,prl,twocolumn,groupedaddress]{revtex4-1}

\usepackage{graphicx}
\usepackage[nointlimits]{amsmath}
\usepackage{amssymb, cancel}

\usepackage{graphics}
\usepackage{dcolumn}

\usepackage{ifpdf}
\usepackage{psfrag}
\usepackage{indentfirst}
\usepackage{bm,verbatim}
\usepackage{color}
\usepackage{cases}
\usepackage{float}
\usepackage{ulem}

\usepackage[colorlinks,linkcolor=blue,urlcolor=black,citecolor=blue]{hyperref}

\usepackage{float}

\usepackage[nointlimits]{amsmath}
\usepackage{amssymb, cancel}
\usepackage{graphicx}
\usepackage{graphics}
\usepackage{dcolumn}
\usepackage{bm}
\usepackage{color}
\usepackage{ifpdf}
\usepackage{psfrag}

\usepackage{lipsum}

\begin{document}

\title {Parametric amplification of continuous variable entangled state for loss-tolerant multi-phase estimation}

\author{\normalsize Sijin Li$^{\mathbf{1}}$ } \email{ Email: sijin.li@polyu.edu.hk}  \author{Wei Wang$^{\mathbf{1}}$}
\email{ Email: wei-iqt.wang@polyu.edu.hk}
\affiliation{$^{1}$ Department of Electrical and Electronic Engineering, The Hong Kong Polytechnic University, Hong Kong 999077, China, \\}
\setlength{\parskip}{0.5ex}

\begin{abstract}
  \noindent 
  Quantum parameter estimation exploits quantum states to achieve estimation sensitivity beyond classical limit.  In continuous variable (CV) regime, squeezed state has been exploited to implement deterministic phase estimation. It is however, often restricted by fragility of quantum states. The quantum phase estimation sensitivity of squeezed state is significantly affected by loss or detection inefficiency, which restrict its applications.  This issue can be solved by using a method of parametric amplification of squeezed state \cite{OPA}. In this work, we implement multi-phase estimation with optical parametric amplification of entanglement generated from squeezed states. We find multi-phase estimation sensitivity is robust against loss or detection inefficiency, where we use two-mode Einstein-Podolsky-Rosen entangled state and four-mode cluster state for analysis.  Our work provides a method for realizing large-scale quantum metrology in real-world applications against loss or detection inefficiency.
\end{abstract}
\maketitle

\subsection*{\normalsize{I. INTRODUCTION}}

\setlength{\parindent}{1em}
   Quantum metrology uses the quantum phenomenon, such as quantum entanglement to estimate unknown physical quantity with precision beyond the classical limit \cite{NPrev}. It has various important applications, including magnetrometry \cite{mag,mag2,mag3}, optical interferometry \cite{interf,interf2}, and atomic clock \cite{clock1,clock2} etc.It can be divided into two aspects, discrete variable (DV) and continuous variable (CV) systems. The DV system uses single photon or entangled multiple photons, described in finite dimensions of Hilbert space, while the CV system uses squeezed state, which is described in infinite dimensions of Hilbert space. In the CV regime, the use of squeezed state can realize deterministic quantum metrology, such as phase estimation \cite{sq1,sq2,sq3}. However, such a squeezed state is sensitive 
to transmission loss and detection inefficiency, which limit its application. A recently proposed method of optical parametric amplification (OPA) of the squeezed state \cite{OPA} can increase both the detection bandwidth and provide the robustness of loss and inefficiency. This method has been used to demonstrate a single phase estimation with OPA of squeezed state \cite{OPA2}, demonstrating quantum enhancement with up to 87$\%$ loss. Also, this technique has been used for demonstrating high bandwidth squeezed state, realizing over 40 GHz bandwidth and 5.2 dB squeezing level \cite{OPA3}.

On the other hand, quantum multi-parameter estimation, which uses multipartite quantum states such as entangled state, can realize parallel measurement of multiple physical quantities. The overall estimation precision can reach the so-called Herseiberg limit $1/N$, where $N$ is the number of entangled states interacting with the measurable quantities, while for the classical system the precision limit is only $1/\sqrt{N}$. Such quantum multi-parameter estimation has attracted much attention recently years in fields including global clock synchronization \cite{clock}, local beam tracking \cite{track}, and quantum imaging \cite{ima}. Quantum multi-parameter estimation is also crucial for quantum network constructin, where the synchronization of parameters such as time \cite{time} and phase \cite{phase} are important for quantum communication throughout network. Such optical quantum multi-parameter estimation of phases is also hindered by fragility of quantum states.

The above mentioned OPA method can also be used in multi-mode system, where each mode of entangled states goes through on OPA. The output entangled state can in general demonstrate loss-tolerance property, which is beneficial for quantum sensing in lossy channels. Recently, this method has been used to demonstrate ultra-fast measurement of two-mode entangled state \cite{fast}. Therefore, this work proposes to use OPA on multi-mode squeezed states, i.e. CV entangled states to realize  loss-tolerant quantum multi-parameter estimation. To begin with, a two-mode entangled state will be used for estimation of two unknown phases, after OPA on each mode of the state. The output state is measured by balanced homodyne detection (BHD) to obtain the amplitude and phase quadratures. The result will be compared with that without OPA, to prove the loss-tolerance property. Then, a four-mode cluster state will be used to realize estimation of four unknown phases.

\subsection*{\normalsize{II. Two mode}}

First, we consider the case of two-mode Einstein-Podolsky-Rosen (EPR) entangled state. We use two single-mode squeezed states:
 \begin{eqnarray}
   &&\hat X_{a1}=\alpha_{1}+ \hat X_{0_{1}}e^{r_{1}}, \hat Y_{a1}=\beta_{1}+ \hat Y_{0_{1}}e^{-r_{1}}\\
   &&\hat X_{a2}=\alpha_{2}+ \hat X_{0_{2}}e^{-r_{2}}, \hat Y_{a2}=\beta_{2}+ \hat Y_{0_{2}}e^{r_{2}},
   \end{eqnarray}
where $\hat X=\hat a+i\hat a^{\dag}$, $\hat Y=-i(\hat a-\hat a^{\dag})$ are amplitude and phase quadratures, $\hat a$ and $\hat a^{\dag}$ are annihilation and creation operators, with vacuum noise, i.e. shot noise limit (SNL) normalized. $\hat X(Y)_{0_{1(2)}}$ is vacuum state quadrature, $ r_{1(2)}$ is squeezing parameter of mode $a1(2)$, $\alpha (\beta)$ is average intensity of amplitude (phase) quadrature.  Next, we generate two-mode EPR entangled state, by interfering two single-mode squeezed states with a 50:50 beam splitter: 
$\hat b_{1}=(\hat a_{1}+\hat a_{2})/\sqrt{2}, \hat b_{2}=(\hat a_{1}-\hat a_{2})/\sqrt{2}$. The EPR state has correlation (anti-correlation) in amplitude (phase), as expressed by their quadrature variances:
 \begin{eqnarray}
   &&\Delta^{2}(\hat X_{b1}-\hat X_{b2})= e^{-2r_{2}}, \Delta^{2}(\hat Y_{b1}+\hat Y_{b2})= e^{-2r_{1}}.
   \end{eqnarray}

\begin{figure}
\begin{center}
\includegraphics[width=1\hsize]{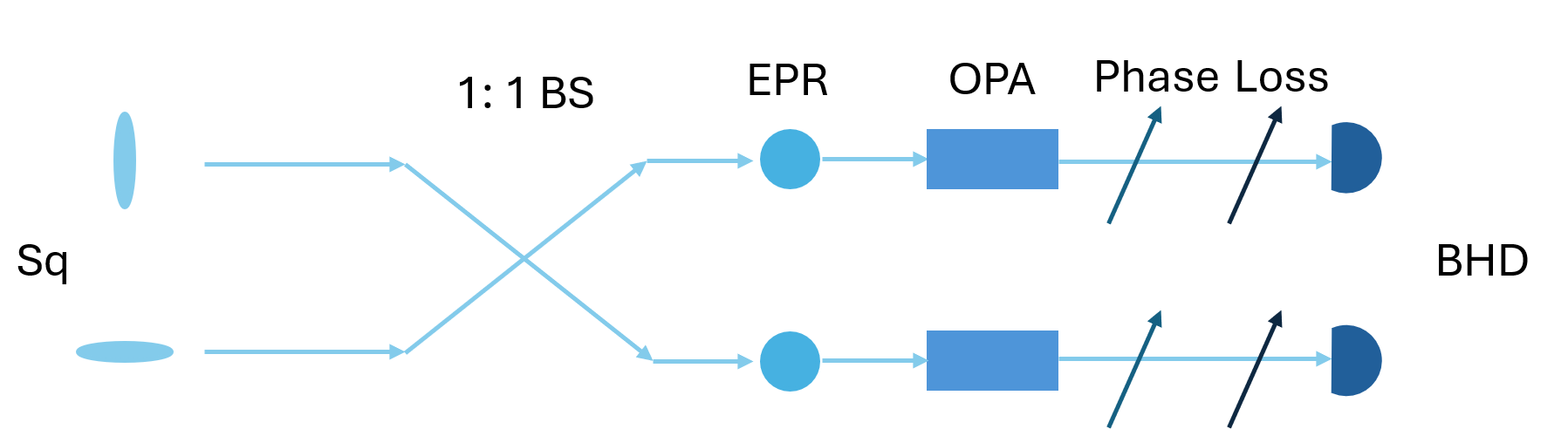}
\caption{Schematic of loss-tolerant phase estimation with two-mode entangled state via OPA. Sq: squeezed state; BS: beam splitter; EPR: Einstein-Podolsky-Rosen two-mode entangled state; BHD: balanced homodyne detection. }
\end{center}
\end{figure}

 The EPR state then passes through two OPAs for amplification. Here, we chosse to amplify phase quadratures of the state:
\begin{eqnarray}
   &&\hat X_{c1}= e^{-r_{3}}\hat X_{b1}, \hat Y_{c1}= e^{r_{3}}\hat Y_{b1},\\
   &&\hat X_{c2}= e^{-r_{4}}\hat X_{b2}, \hat Y_{c2}= e^{r_{4}}\hat Y_{b2},
   \end{eqnarray}
where $r_{3}, r_{4}$ are squeezing parameter of two OPAs. We can see that the EPR correlation of the two modes remains as we set $r_{3}= r_{4}$. Next, two unknown phases are imposed on the two modes, i.e. $e^{i \theta_{1}}  \hat c_{1}, e^{i \theta_{2}}  \hat c_{2}$. Last we perform BHD on the two modes:
\begin{eqnarray}
   &&I_{1}= I_{LO}[\hat X_{c1}cos(\theta_{1}+\phi_{1})+\hat Y_{c1}sin(\theta_{1}+\phi_{1})],\\
   &&I_{2}= I_{LO}[\hat X_{c2}cos(\theta_{2}+\phi_{2})+\hat Y_{c2}sin(\theta_{2}+\phi_{2})],
   \end{eqnarray}
where $\phi$ is relatvie between signal and LO, and can be locked via phase locking. Thus we can measure and estimate two unknown phases $\theta_{1},  \theta_{2}$ by the above joint BHDs. We then consider the loss, including transmission loss and detection inefficiency, which can be modelled as a vacuum state introduced via a beam spliter: $\hat a=\eta a_{0}+(1-\eta) \nu_{0}$, where $\nu_{0}$ is vacuum state, $1-\eta$ is loss. Thus the two BHDs can be rewritten as:
\begin{small}
\begin{eqnarray}
   I_{1(2)}\approx I_{LO}[(\eta_{1(2)}\hat Y_{c1(2)}+(1-\eta_{1(2)})\hat Y_{\nu 1(2)})sin(\theta_{1(2)}+\phi_{1(2)})],
\end{eqnarray}
\end{small}
where the $\hat X$ terms are removed since the OPAs amplifies $\hat Y$ and deamplifies $\hat X$, i.e. $\hat X \ll \hat Y$. Next, the phase estimation sensitivity can be quantified by \cite{distribute}:
\begin{eqnarray}
  \sigma_{\theta_{1(2)}}=\frac{\sqrt{\Delta ^{2}(P)}}{\partial <P>/\partial \theta_{1(2)}},
   \end{eqnarray}
where $ P$ stands for estimator of phase, $\Delta ^{2}(P)$ is variance, $<P>$ is expectation value, $\partial (P)/\partial \theta_{1(2)}$ is slope of $P$ versus $\theta_{1(2)}$. Obviously, this sensitivity can also be quantified by using joint BHDs of $I_{1} +I_{2}$ (corresponding to phase correlations of EPR state):
\begin{eqnarray}
  \sigma_{\theta_{1(2)}}=\frac{\sqrt{\Delta ^{2}(I_{1}\pm I_{2})}}{\partial <I_{1}\pm I_{2}>/\partial \theta_{1(2)}},
   \end{eqnarray}
taking into above equations, we can calculate the sensitity:
\begin{small}
\begin{equation}
\begin{split}
  \sigma_{\theta_{1(2)}}=&\sqrt{(\eta_{1}^{2}+\eta_{2}^{2})(e^{-2r_{1}}+e^{2r_{2}})+2\eta_{1}\eta_{2}(e^{-2r_{1}}-e^{2r_{2}})} \\ &\sqrt{+[(1-\eta_{1}^{2})+(1-\eta_{2}^{2})]e^{-2r_{3}}}/ \frac{\eta_{1}\theta_{1(2)}|\beta_{1}\pm\beta_{2}|}{\sqrt{2}} ,\\
\end{split}
\end{equation}
\end{small}

\begin{figure}
\begin{center}
\includegraphics[width=1\hsize]{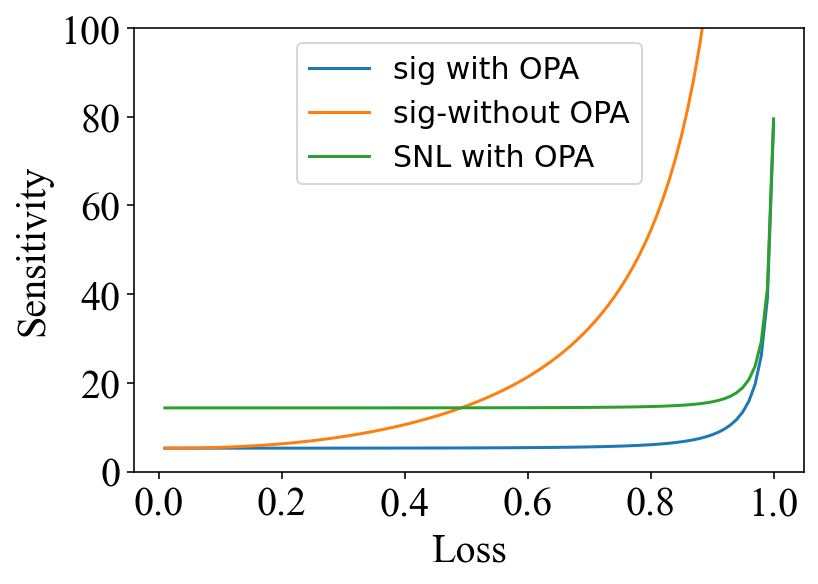}
\caption{Average phase estimation sensitivity of two phases as a function of loss. The losses in every beam are set to the same value. $r_{1}=1$, corresponding to initial squeezing of about 8 dB. $r_{1}^{'}=4.6$, corresponding to OPA gain of 100. $\beta_{1}=1$, $\beta_{2}5$, $\theta_{1}=1.5^{\circ}$}.
\end{center}
\end{figure}

where we have assumed small phase, i.e. $\theta_{1(2)} \ll 1, cos\theta_{1(2)} \approx 1$, and set $\phi=90^{\circ}$, $r_{3}=r_{4}$. Since the EPR state is symmetric, it is the same for the situation of amplifying the amplitude quadrature. Thus we only use the phase quadrature correlation to calculate the sensitivity of estimating two phases. The results of average sensitivity of estimating two phases are shown in Fig 2, where the initial squeezing is set to 8 dB, with $\eta_{1}=\eta_{2}$ for simplicity of calculation. It is seen in Fig. 2 that with OPA, the sensitivity is constant under loss, even when loss reaches 95$\%$. However, for sensitivity without OPA, it degrades rapidly with loss, corresponding to the orange line. This clearly demonstrates the advantage of OPA for realizing loss-tolerant phase estimation. In addition, the SNL is calculated by setting $r_{1}=0$, with OPA present, corresponding to the green line. The quantum advantage is demonstrated with 8 dB of initial squeezing in almost all losses.

  \subsection*{\normalsize{III. Four mode}}
Next, we extend to the situation with four modes, exploiting the CV cluster state \cite{cluster1}. The cluster can be described by the quadratures:
\begin{eqnarray}
  (\hat{Y_{a}}-\sum_{b \in N_{a}}\hat{X_{b}}) \rightarrow 0
   \end{eqnarray}
 Here, modes b are the nearest neighbors of mode a. Here we exploit the method of beam splitter along with the off-line squeezed state for the generation of the cluster state \cite{cluster2}. Various shapes of cluster states can be generated with this method, including T-shaped, square-shaped, and linear cluster states \cite{cluster3}.

Figure 3 shows the schematic of loss-tolerant phase estimation based on four-mode cluster state. To begin with, four initial single-modes states squeezed in phase quadrature are prepared: 
 \begin{eqnarray}
  &&\hat X_{a(i)}=\alpha _{i}+\hat X_{0(i)}e^{r_{i}},\\
 &&\hat Y_{a(i)}=\beta _{i}+\hat Y_{0(i)}e^{-r_{i}}, (i=1,2,3,4)
  \end{eqnarray}
$X_{0(i)},Y_{0(i)}$ are vacuum state quadratures. Then the cluster state can be generated by applying a specific unitary matrix on the four states: $\hat b_{i}=\sum_{j} U_{ij}\hat a_{j}$. The unitary matrix changes as we choose a different cluster state shape. Here we consider the square-shaped cluster state, whose unitary matrix is \cite{cluster3}:
\begin{equation}
U=
\left(
\begin{array}{cccc}
 -\frac{1}{\sqrt{2}} & -\frac{1}{\sqrt{10}} & -\frac{2i}{\sqrt{10}} & 0 \\
 \frac{1}{\sqrt{2}} & -\frac{1}{\sqrt{10}} & -\frac{2i}{\sqrt{10}} & 0 \\
 0 & -\frac{2i}{\sqrt{10}} & -\frac{1}{\sqrt{10}} & -\frac{1}{\sqrt{2}} \\
  0 & -\frac{2i}{\sqrt{10}} & -\frac{1}{\sqrt{10}} & \frac{1}{\sqrt{2}} 
\end{array}
\right)
\end{equation}

\begin{figure}
\begin{center}
\includegraphics[width=1\hsize]{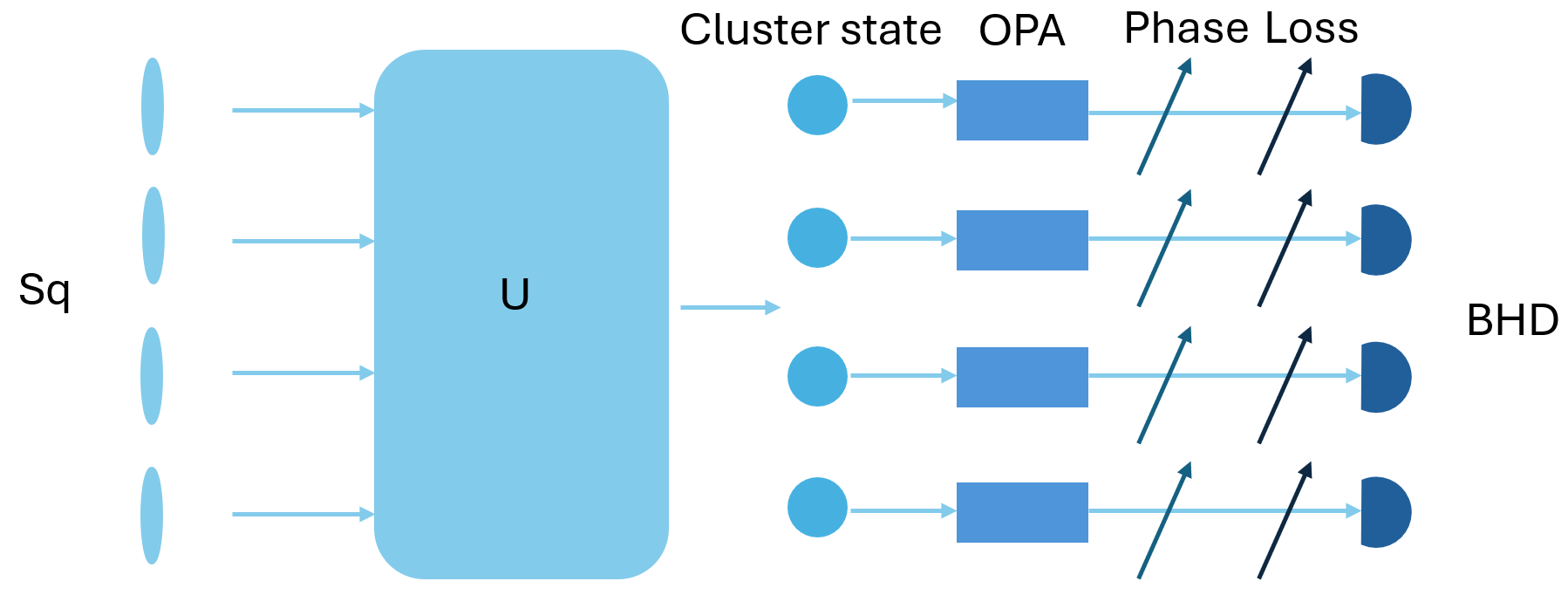}
\caption{Schematic of loss-tolerant phase estimation with four-mode entangled state via OPA. Sq: squeezed state; U: unitary matrix for construction of cluster state; BHD: balanced homodyne detection. }
\end{center}
\end{figure}

The output cluster state has quantum correlations:
\begin{eqnarray}
  \hat Y_{b1}-\hat X_{b3}-\hat X_{b4}=-\frac{1}{\sqrt{2}}e^{-r_{1}}\hat Y_{a(1)}-\frac{\sqrt{5}}{\sqrt{2}}e^{-r_{2}}\hat Y_{a(2)},\\
  \hat Y_{b2}-\hat X_{b3}-\hat X_{b4}=\frac{1}{\sqrt{2}}e^{-r_{1}}\hat Y_{a(1)}-\frac{\sqrt{5}}{\sqrt{2}}e^{-r_{2}}\hat Y_{a(2)},\\
  \hat Y_{b3}-\hat X_{b1}-\hat X_{b2}=-\frac{\sqrt{5}}{\sqrt{2}}e^{-r_{3}}\hat Y_{a(3)}-\frac{1}{\sqrt{2}}e^{-r_{4}}\hat Y_{a(4)},\\
  \hat Y_{b4}-\hat X_{b1}-\hat X_{b2}=-\frac{\sqrt{5}}{\sqrt{2}}e^{-r_{3}}\hat Y_{a(3)}+\frac{1}{\sqrt{2}}e^{-r_{4}}\hat Y_{a(4)},
  \end{eqnarray}
next, the output four-mode  cluster state needs to be amplified by OPAs. However, since the above correlations are different combinations of various modes among the cluster state, OPA angles need to be adjusted as different correlations are used. Meanwhile, the four-mode cluster state can estimate four unknown phases. For each phase estimation, a joint BHD on the cluster will be conducted. Thus, four times of joint BHDs will be performed in order to estimate four unknown phases. As an example, we first use the correlation of Eq (16) for estimation of phase in mode 1.  Obviously, phase quadrature of mode 1 and amplitude quadratures of mode 3 and 4 will be amplified by OPA: $\hat Y_{c1}=e^{r_{1}^{'}}\hat Y_{b1}, \hat X_{c1}=e^{-r_{1}^{'}}\hat X_{b1}, \hat Y_{c3(4)}=e^{-r_{3(4)}^{'}}\hat Y_{b3(4)}, \hat X_{c3(4)}=e^{r_{3(4)}^{'}}\hat X_{b3(4)}$. Then the joint BHD on these three modes can be performed. For each BHD, we use an approximation to remove the de-amplification term, e.g.:
\begin{small}
\begin{eqnarray}
  &I_{1}=I_{LO}[\eta_{1}\hat X_{c1}cos(\theta_{1}+\phi_{1})+(1-\eta_{1})\hat X_{vac1}cos(\theta_{1} \nonumber \\ \nonumber
  &+\phi_{1})+\eta_{1}\hat Y_{c1}sin(\theta_{1}+\phi_{1})+(1-\eta_{1})\hat Y_{c1}sin(\theta_{1}+\phi_{1})] \\
  & \approx I_{LO}[\eta_{1}\hat Y_{c1}sin(\theta_{1}+\phi_{1})+(1-\eta_{1})\hat Y_{c1}sin(\theta_{1}+\phi_{1})],
  \end{eqnarray}
\end{small}
where $1-\eta_{1}$ is loss in this mode, $\theta_{1}$ is the unknonw phase, $\phi_{1}$ is the BHD phase. Then four different joint BHDs will be performed based on the quantum correlations of Eqs (16)-(19). The phases estimation sensitivity of four phases will be separately obtained from four different joint BHDs. To begin with, the joint BHD from Eq. (16) is calculated for sensitivity of phase 1:


\begin{small}
\begin{eqnarray}
  \sigma^{'}_{\theta_{1}}=&\frac{\sqrt{\Delta ^{2}(I_{1}- I_{3}-I_{4})}}{\partial <I_{1}- I_{3}-I_{4}>/\partial \theta_{1}}  \nonumber\\
  =&\sqrt{e^{-2r_{1}}[\frac{1}{2}\eta_{1}^{2}+\frac{1}{10}(\eta_{1}+2\eta_{3}+2\eta_{4})^{2}]+} \nonumber \\
  &\sqrt{e^{2r_{1}}[\frac{1}{10}(2\eta_{1}-\eta_{3}-\eta_{4})^{2}+\frac{1}{2}(\eta_{3}-\eta_{4})^{2}]} \nonumber \\
  &\sqrt{+e^{-2r_{1}^{'}}[(1-\eta_{1})^{2}+(1-\eta_{3})^{2}+(1-\eta_{4})^{2}]} \nonumber \\
 & /[\theta_{1}\eta_{1}(\frac{1}{\sqrt{2}}\beta_{1}+\frac{1}{\sqrt{10}}\beta_{2}+\frac{2}{\sqrt{10}}\alpha_{3})],
  \end{eqnarray}
\end{small}
where we have also made approximations, i.e. $\theta_{1} \ll 1, cos\theta_{1} \approx 1$. The initial squeezing values are set to the same value $r_{1}$, and OPA gains are also set to the same $r_{1}^{'}$.
\begin{figure}
\begin{center}
\includegraphics[width=1\hsize]{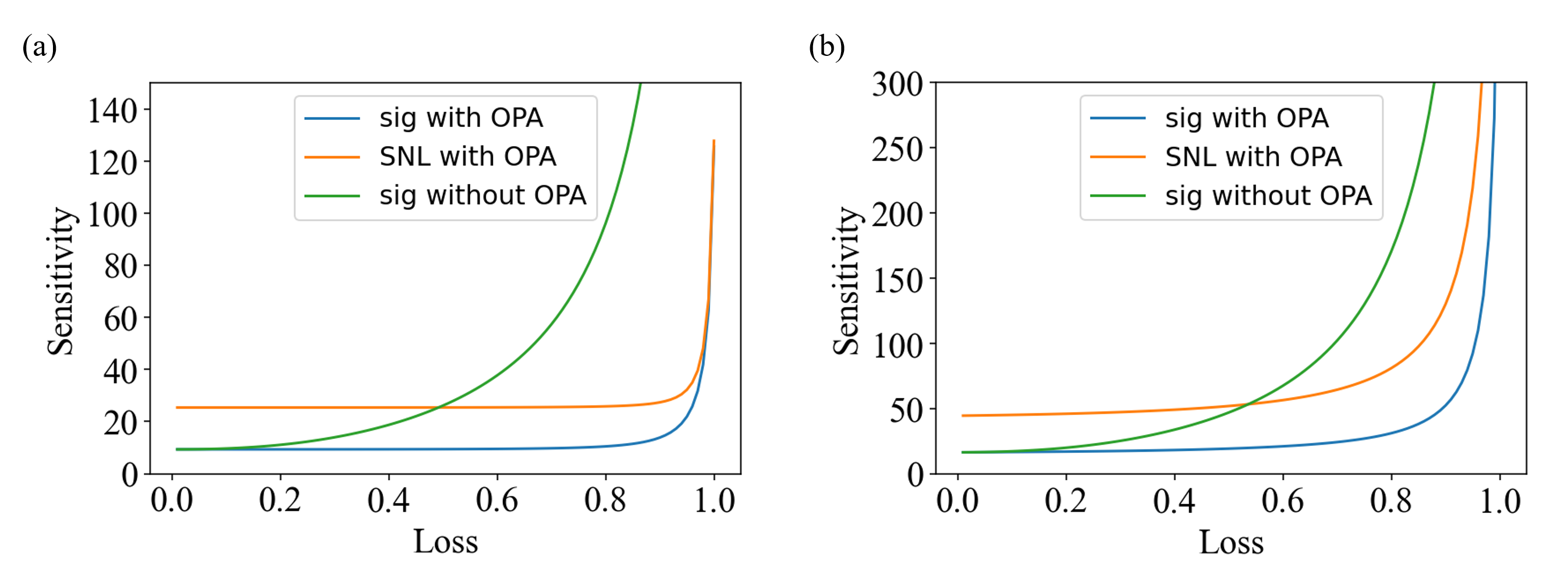}
\caption{(a)Estimation sensitivity of phase 1 as a function of loss. (b)Average phase estimation sensitivity of four phases as a function of loss. The losses in every beam are set to the same value. $r_{1}=1$, corresponding to initial squeezing of about 8 dB. $r_{1}^{'}=3$, corresponding to OPA gain of 20. $\beta_{1}=1$, $\beta_{2}=\alpha_{3}=2$, $\theta_{1}=1.5^{\circ}$}.
\end{center}
\end{figure}
We then calculate the sensitivity as a function of loss, as shown in Fig. 4(a). We have set all losses to the same value. It can be seen that sensitivity remain almost constant for loss under 90$\%$. For comparison, we also plotted sensitivity of SNL after OPA, i.e. $r_{1}=0$, and the sensitivity when OPA is removied, i.e. $r_{1}^{'}=0$. The quantum enhancement is clearly demonstrated compared with SNL, and the sensitivity without OPA degrades as loss is increased, and is larger than SNL when loss is over 50$\%$, which is in agreement with the result of Fig. 2. Next, the sensitivity of other phase estimation is calculated. The sensitivity from Eq. (17) is:
\begin{small}
\begin{eqnarray}
  \sigma^{'}_{\theta_{2}}=&\frac{\sqrt{\Delta ^{2}(I_{2}- I_{3}-I_{4})}}{\partial <I_{2}- I_{3}-I_{4}>/\partial \theta_{2}} \nonumber \\
  =&\sqrt{e^{-2r_{1}}[\frac{1}{2}\eta_{2}^{2}+\frac{1}{10}(\eta_{2}+2\eta_{3}+2\eta_{4})^{2}]+} \nonumber \\
  &\sqrt{e^{2r_{1}}[\frac{1}{10}(2\eta_{2}-\eta_{3}-\eta_{4})^{2}+\frac{1}{2}(\eta_{3}-\eta_{4})^{2}]} \nonumber \\
  &\sqrt{+e^{-2r_{1}^{'}}[(1-\eta_{1})^{2}+(1-\eta_{2})^{2}+(1-\eta_{3})^{2}]} \nonumber \\
  &/[\theta_{2}\eta_{2}(\frac{1}{\sqrt{2}}\beta_{1}-\frac{1}{\sqrt{10}}\beta_{2}-\frac{2}{\sqrt{10}}\alpha_{3})],
  \end{eqnarray}
\end{small}
where the initial squeezing and OPA gain are all set to the same value as in phase 1. The sensitivity of phase 3 and 4 can thus be calculate from Eqs. (18) and (19):
\begin{small}
\begin{eqnarray}
  \sigma^{'}_{\theta_{3}}=&\frac{\sqrt{\Delta ^{2}(I_{3}- I_{1}-I_{2})}}{\partial <I_{3}- I_{1}-I_{2}>/\partial \theta_{3}} \nonumber \\
  =&\sqrt{e^{-2r_{1}}[\frac{1}{2}\eta_{3}^{2}+\frac{1}{10}(2\eta_{1}+2\eta_{2}+\eta_{3})^{2}]+} \nonumber \\
  &\sqrt{e^{2r_{1}}[\frac{1}{10}(\eta_{1}+\eta_{2}-2\eta_{3})^{2}+\frac{1}{2}(\eta_{1}-\eta_{2})^{2}]} \nonumber \\
  &\sqrt{+e^{-2r_{1}^{'}}[(1-\eta_{1})^{2}+(1-\eta_{2})^{2}+(1-\eta_{3})^{2}]} \nonumber \\
  &/[\theta_{3}\eta_{3}(\frac{2}{\sqrt{10}}\alpha_{2}+\frac{1}{\sqrt{10}}\beta_{2}+\frac{1}{\sqrt{2}}\beta_{4})],
  \end{eqnarray}
\end{small}
and:
\begin{small}
\begin{eqnarray}
  \sigma^{'}_{\theta_{4}}=&\frac{\sqrt{\Delta ^{2}(I_{4}- I_{1}-I_{2})}}{\partial <I_{4}- I_{1}-I_{2}>/\partial \theta_{4}} \nonumber \\
  =&\sqrt{e^{-2r_{1}}[\frac{1}{2}\eta_{4}^{2}+\frac{1}{10}(2\eta_{1}+2\eta_{2}+\eta_{4})^{2}]+} \nonumber \\
  &\sqrt{e^{2r_{1}}[\frac{1}{10}(\eta_{1}+\eta_{2}-2\eta_{4})^{2}+\frac{1}{2}(\eta_{1}-\eta_{2})^{2}]} \nonumber \\
  &\sqrt{+e^{-2r_{1}^{'}}[(1-\eta_{1})^{2}+(1-\eta_{2})^{2}+(1-\eta_{4})^{2}]} \nonumber \\
  &/[\theta_{4}\eta_{4}(\frac{2}{\sqrt{10}}\alpha_{2}+\frac{1}{\sqrt{10}}\beta_{2}-\frac{1}{\sqrt{2}}\beta_{4})].
  \end{eqnarray}
\end{small}
The result of the average estimation sensitivity of four phases, $\sigma_{ave}$, is shown in Fig. 4 (b). It can be seen that the result is similar to that of Fig. 4(a). The slight difference is due to the asymmetric structure of the cluster state, which makes the estimation sensitivity of phase in each mode different.

We also further study the phase estimation sensitivity with asymmetric loss situations, as shown in Fig. 5. Fig. 5(a) shows the average phase estimation sensitivity of 4 phases as function of losses in modes 1 and 2, with $\eta_{3}=\eta_{4}=0.5$. It can be seen that the sensitivity with OPA is still tolerant to two losses. Also, the sensitivity with OPA has an optimal point. The sensitivity without OPA degrades rapidly with losses and then becomes larger than SNL. The estimation sensitivity with different combinations of losses in two modes are shown in Figs 5(b)-(d), with the remaining two modes all set to 0.5. The sensitivities all have the similar trends. These figures demonstrate the multi-phase estimation sensitivities with different losses combinations, and can be helpful for experimental realizations of loss-tolerant multi-parameter estimation with cluster state.

\begin{figure}
\begin{center}
\includegraphics[width=1\hsize]{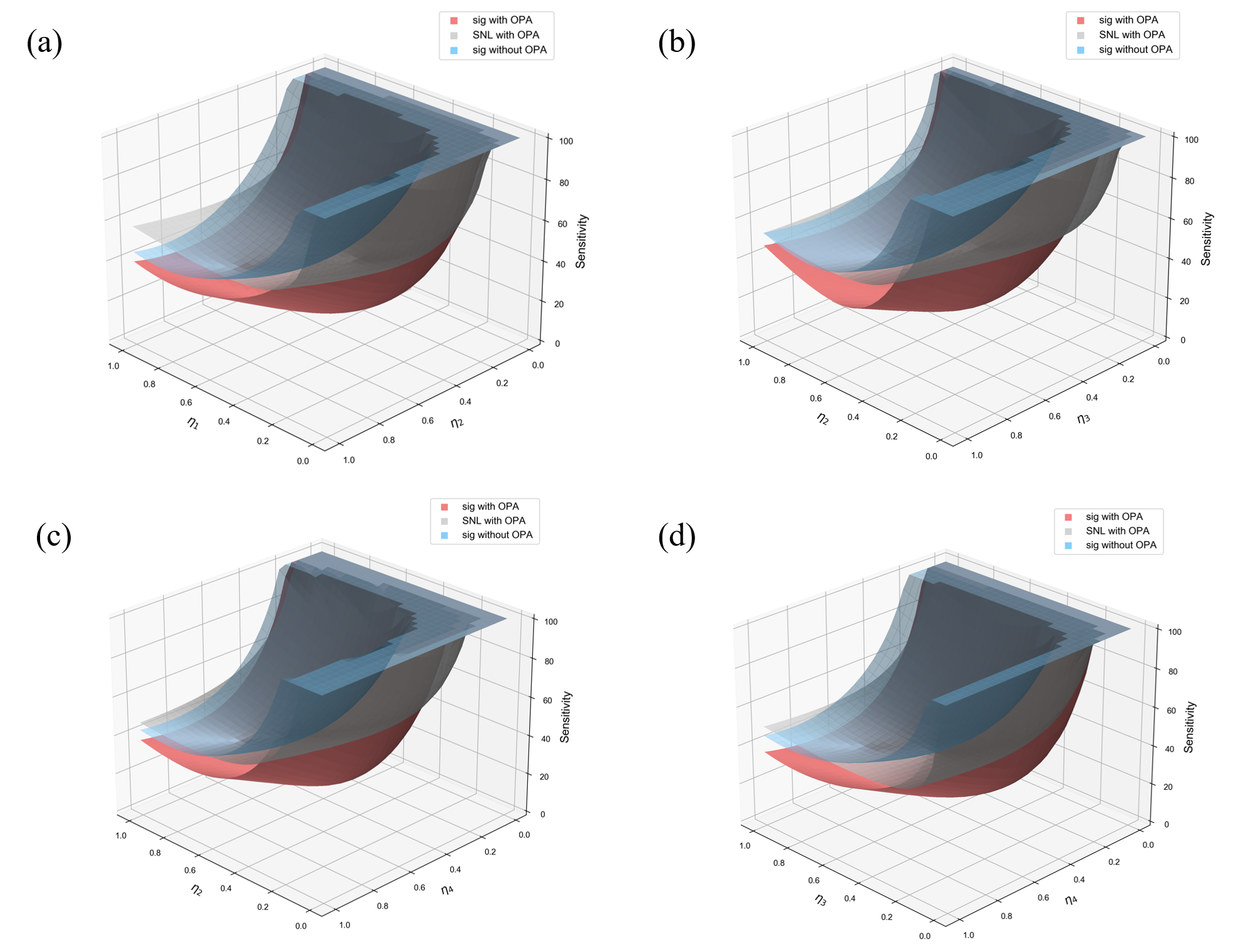}
\caption{Average phase estimation sensitivity of 4 phases as a function of losses in: (a)modes 1 and 2, with $\eta_{3}=\eta_{4}=0.5$; (b)modes 2 and 3, with $\eta_{1}=\eta_{4}=0.5$; (c)modes 2 and 4, with $\eta_{3}=\eta_{4}=0.5$; (d)modes 3 and 4, with $\eta_{3}=\eta_{4}=0.5$. Other parameters: $r_{1}=1$, corresponding to initial squeezing of about 8 dB. $r_{1}^{'}=3$, corresponding to OPA gain of 20. $\beta_{1}=1$, $\beta_{2}=\alpha_{3}=2$, $\alpha_{2}=1$, $\beta_{4}=3$, $\theta_{1}=1.5^{\circ}$. (a)}
\end{center}
\end{figure}

\subsection*{\normalsize{IV. CONCLUSION}}

  We derived the multi-phase estimation sensitivities via OPA on multi-mode entangled states, including EPR state and four-mode cluster state. The sensitivities with OPA are compared with SNL and sensitivities without OPA. By introducing losses on the entangled state, the loss-tolerance properties are verified as OPA is present. For the four-mode case, two losses are varied simultaneously to further study the loss-tolerance property. These studies provide a method of realizing practical quantum metrology under lossy channels using CV entangled state, and lay foundation for subsequent experimental implementations.

\subsection*{\normalsize{ ACKNOWLEDGEMENTS}}

\normalsize{This work was funded by .}\\

\end{document}